%% file: Template.tex
\documentclass{article}
\usepackage{spconf,amsmath,graphicx}

\usepackage{multirow}
\usepackage{makecell}
\usepackage{color,soul} 
\usepackage{enumitem}
\usepackage{algorithm}
\usepackage{algorithmic}
\usepackage{graphicx}
\usepackage{wrapfig,lipsum,booktabs}
\usepackage{enumitem}
\usepackage{url}
\usepackage{cite}
\usepackage{subcaption} 
\usepackage{authblk}
\usepackage{fancyhdr}

\usepackage{amsmath}
\usepackage{amssymb}
\usepackage{mathtools}
\usepackage{amsthm}

\usepackage{tikz}
\usepackage{textcomp}
\usepackage{hyperref}
\usepackage{lipsum}

\newcommand\copyrighttext{%
  \footnotesize \textcopyright 2023 IEEE. Personal use of this material is permitted. Permission from IEEE must be obtained for all other uses, in any current or future media, including reprinting/republishing this material for advertising or promotional purposes, creating new collective works, for resale or redistribution to servers or lists, or reuse of any copyrighted component of this work in other works. DOI: 
  }
\newcommand\copyrightnoticeieee{%
\begin{tikzpicture}[remember picture,overlay]
\node[anchor=south,yshift=10pt] at (current page.south) {\fbox{\parbox{\dimexpr\textwidth-\fboxsep-\fboxrule\relax}{\copyrighttext}}};
\end{tikzpicture}%
}

\input{defs.tex}

\title{To Wake-up or Not to Wake-up: Reducing Keyword False Alarm by Successive Refinement}
\name{Yashas Malur Saidutta*, Rakshith Sharma Srinivasa*, Ching-Hua Lee, \\  Chouchang Yang, Yilin Shen, Hongxia Jin \thanks{* Equal contribution.}}
\address{Samsung Research America, Mountain View, CA, USA}
\usepackage{background}
\SetBgContents{\color{black}Published as a conference paper at ICASSP 2023}
\SetBgScale{1}
\SetBgAngle{0}
\SetBgPosition{current page.north east}
\SetBgHshift{-11cm}
\SetBgVshift{-1cm}

\begin{document}
\ninept
\maketitle
\copyrightnoticeieee
\begin{abstract}
Keyword spotting systems continuously process audio streams to detect keywords. One of the most challenging tasks in designing such systems is to reduce \emph{False Alarm (FA)} which happens when the system falsely registers a keyword despite the keyword not being uttered. In this paper, we propose a simple yet elegant solution to this problem that follows from the law of total probability. We show that existing deep keyword spotting mechanisms can be improved by \emph{Successive Refinement}, where the system first classifies whether the input audio is speech or not, followed by whether the input is keyword-like or not, and finally classifies which keyword was uttered. We show across multiple models with size ranging from 13K parameters to 2.41M parameters, the successive refinement technique reduces FA by up to a factor of 8 on in-domain held-out FA data, and up to a factor of 7 on out-of-domain (OOD) FA data. Further, our proposed approach is ``plug-and-play" and can be applied to any deep keyword spotting model. 
\end{abstract}

\input{Contents/Introduction}
\input{Contents/Previous_work}

\input{Contents/Proposed_method}
\input{Contents/Experiments}

\input{Contents/Conclusion}

\footnotesize
\bibliographystyle{IEEEbib}
\bibliography{strings,refs}

\end{document}

%% file: defs.tex
\renewcommand{\P}[1]{\operatorname{p}\left(#1\right)}

\newcommand{\calL}{\mathcal{L}}
\newcommand{\defeq}{\vcentcolon=}
\DeclareMathOperator*{\argmax}{arg\,max}

%% file: Contents/Introduction.tex
\section{Introduction}
\label{sec:introduction}
From mobile devices to home appliances, keyword spotting (KWS) forms a cornerstone of human device interaction \cite{haeb2019speech}. KWS systems continuously process audio streams to detect keywords. However, during a majority of the time, KWS systems only process audio that contains non-keyword speech or non-speech (also referred to as noise), since keywords are uttered only sparsely. Consider an always-on KWS system with a false alarm (FA) rate of $5\%$ processing one second of audio every 0.1 second (due to overlap with previous audio frames). This results in a FA count of \textit{1800} per hour. Even more importantly, upon detecting a keyword KWS systems are used to trigger other larger cloud-based systems like the Automatic Speech Recognition (ASR). Thus, an FA in the KWS system could lead to privacy concerns due to unnecessary recording and uploading of user audio to the cloud. Finally, triggering of other downstream tasks leads to unwanted power consumption, which is undesirable in battery-powered devices like smart phones and smart watches.

\begin{figure}[t]
\centering
    \begin{subfigure}{0.2\textwidth}
    \centering
        \includegraphics[ width=0.35\linewidth]{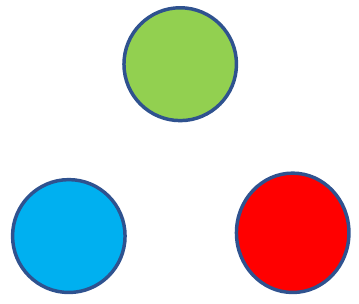}
        \caption{Prior art point-of-view}
        \label{fig:prior_art_venn}
    \end{subfigure}%
    \begin{subfigure}{0.25\textwidth}
    \centering
        \includegraphics[width=0.5\linewidth]{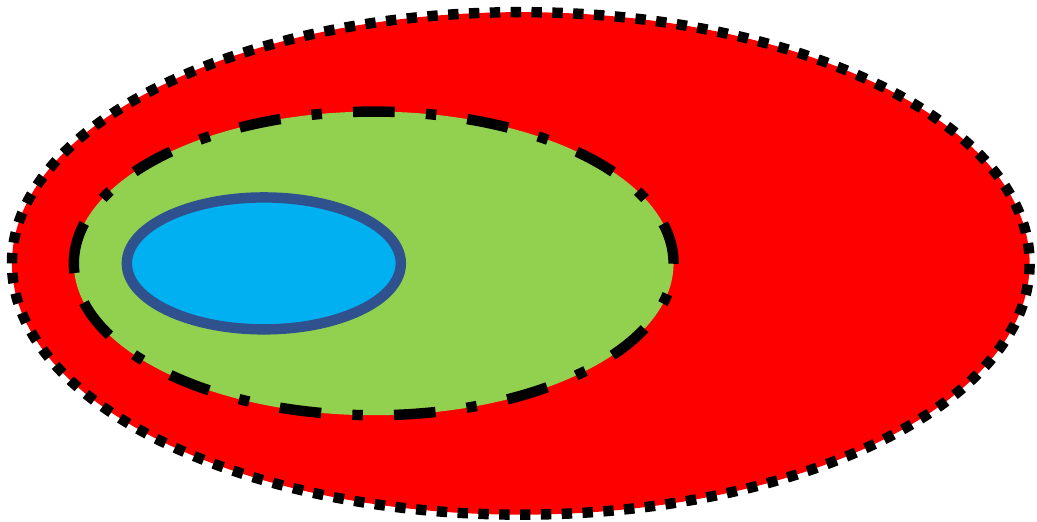}
        \caption{Proposed point-of-view}
        \label{fig:our_mech_venn}
    \end{subfigure}
\caption{Here, red represents non-speech, green represents non-keyword speech, and blue represents keyword(s). (a) Prior works look at these sets as mutually exclusive classes \textit{equally far-apart}. (b) In the proposed point-of-view, all audio are bounded by the dotted line. The speech component forms a smaller subset within all audio. The keyword(s) form a smaller subset within the speech set.}
\label{fig:venn_diagrams}
\vspace{-5mm}
\end{figure}

Prior KWS works employing deep neural networks (DNNs), namely, deep KWS systems, have obtained impressive results \cite{lopez2021deep}. However, they treat non-keyword speech and non-speech audio as two mutually exclusive classes that are equally distant from the set of keywords. Fig. \ref{fig:venn_diagrams} (a) showcases this point-of-view in a single keyword system. However, in reality, speech forms a subset in the set of all audio, and keywords form a subset within speech; as shown in Fig. \ref{fig:venn_diagrams} (b). The proposed point-of-view allows us to leverage the fact that keywords share more similarities with non-keyword speech, owing to the fact that they are both speech. A similar argument can be made for keywords. Keywords are spoken as stand-alone commands and thus have a characteristic rise and fall in energy associated with the utterance. On the contrary, general speech has a more stable energy profile. 

Based on the above insights, we propose a \emph{Successive Refinement} based deep KWS classifier. The classifier first classifies whether the input audio is speech or not. If it is speech, it classifies whether the speech is keyword-like or non-keyword speech. Finally, if it is deemed keyword-like, the final classifier identifies which keyword was uttered. Crucially, this methodology does not attempt to deploy three separate classifier models. Instead, we re-use internal representations of a backbone model for the three different classification branches, where each branch is implemented by replicating the last one or at most two layers of the model.
Such a hierarchical supervision leads to richer learning, and allows for training models with reduced FA. We summarize our contributions below:
\begin{itemize}[leftmargin=4mm]
    \itemsep0em
    \item We develop a generalized method to reduce FA rate in deep KWS models which can be used in any chosen architecture, with negligible increase in memory requirement and computation.
    \item On state-of-the-art (SOTA) deep KWS models, 
    incorporating \emph{Successive Refinement} leads to a reduction in FA rates by up to a factor of 8 on held-out FA data.
    \item \emph{Successive Refinement} shows consistent improvement on OOD FA data by up to a factor of 7. The performance on OOD is crucial because it is impossible to expose the model to all possible audio data during training.    
\end{itemize}
\vspace{-5mm}

\begin{figure*}
\centering
\includegraphics[width=\linewidth]{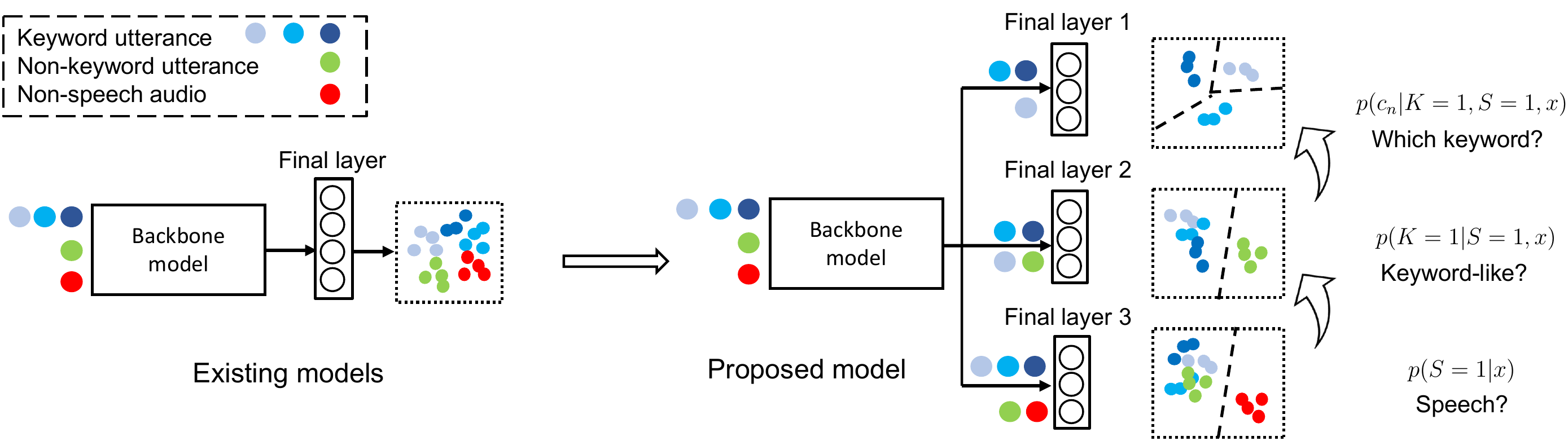}
\caption{System Overview: In this paper, we present a new approach to mitigating FA in KWS. Our proposed approach treats the keyword detection problem as a multi-stage classification problem and uses successive refinement to filter out non-keyword audio.}
\label{fig:overview}
\vspace{-5mm}
\end{figure*}

    
    

%% file: Contents/Previous_work.tex
\section{Related Works}


There have been many KWS works using deep learning techniques for achieving superior performance to conventional model-based approaches, e.g., \cite{sun2016max, arik2017convolutional, sainath2015convolutional, jimmy2018deep, Choi2019temporal, li2020small, xu2020depthwise, kim2021broadcasted, mengjun2019effective, rybakov2020streaming, berg21_interspeech, Bai2019time, de2018neural, Xi2019small}. 
Works in deep KWS have attempted to learn parameterized feature extraction from raw audio \cite{mittermaier2020small} or learn the architecture itself \cite{zhang2021autokws, mo2020neural}. Another important aspect of deep KWS systems is the model size. Since KWS models are usually deployed on devices and are continuously processing an audio stream, they need to have low memory and computational footprint. There have been several papers that focus on building models that meet this criteria for KWS \cite{bruno2018small, Myer2018efficient, kim2021broadcasted}. For a comprehensive review of deep KWS systems we refer the reader to \cite{lopez2021deep}.

Different from above works, another crucial aspect for deep KWS is FA reduction. It is important to have a low FA rate due to privacy and power consumption concerns, as described in Section \ref{sec:introduction}. An efficient FA mitigation system is designed in \cite{garg21_interspeech}, but they require post-utterance audio to gain contextual information. Recently in \cite{wang2020wake}, the authors indicated that a good non-speech hidden Markov model is crucial for good performance of a wakeword system. In this paper, our motivation is similar, i.e., we leverage a good Speech vs Non-Speech and a good Keyword vs Non-Keyword deep classifiers for better performance. The proposed method is complementary to the above works and can be applied to enhance any of them.
\vspace{-2mm}

%% file: Contents/Proposed_method.tex
\section{Reducing FA using successive refinement}

The core idea of our method is based on the following observation: existing keyword detection models treat keyword utterances, non-keyword utterances and non-speech audio simply as belonging to different classes. Based on this assumption, existing models treat keyword detection as a multi-class classification problem. However, these different kinds of audio have a more natural hierarchical structure to them, as described in Section \ref{sec:introduction}. 

The hierarchical structure described above can be quantified using the law of total probability as shown below. Let us consider a keyword detection system with $N$ words in the vocabulary. Let $c_n$ denote the $n$th keyword. First, we define a random variable $S$ to be $1$ when the audio frame $x$ contains speech and $0$ otherwise. We also define another random variable $K$ to be $1$ when $x$ contains keyword-speech and $0$ when it contains non-keyword speech. Then, for a given input audio $x$, we have 
\vspace{-2mm}
\begin{align}
    \mkern-20mu p(c_n|x)&=\text{Probability of } x \text{ being keyword } c_n \\
    =\sum_{k,s\in\{0,1\}}&p_{C|K,S,X}(c_n|k,s,x)p_{K|S,X}(k|s,x)p_{S|X}(s|x) \label{eq: chain rule 1} \\
    =p(c_n|k&=1,s=1,x)p(k=1|s=1,x)p(s=1|x) \label{eq: chain rule 2},
\end{align}
where (\ref{eq: chain rule 1}) follows from the hierarchical nature of the audio classes (Fig.~\ref{fig:venn_diagrams} (b)) and (\ref{eq: chain rule 2}) follows because $p(c_n)=0$ if $K=0$ or $S = 0$. 

We use the above hierarchical structure of the output probability to design our model. A schematic overview of the proposed system to perform keyword detection is shown in Fig. \ref{fig:overview}. As seen in the figure, our architecture consists of a backbone neural network followed by three branches: \textit{keyword classification, keyword branch } and \textit{speech branch}. All branches consist of at most two layers, with sigmoid activation for the keyword and speech branch, and softmax activation for the keyword classification branch. The speech branch is trained to classify a given input as human speech or otherwise, and the keyword branch is trained to classify a given input as one of the keyword, or otherwise. Finally, the keyword classification branch is trained to classify which keyword is present in the given input. The way these branches are used during training and inference differ, as outlined below.

\vspace{0.15cm}

\noindent\textbf{Training:} During training, we model the data flow through the network using \eqref{eq: chain rule 2}. The backbone model first generates intermediate embeddings for the entire mini-batch. Then, embeddings corresponding to only those inputs containing keywords are passed through the keyword branch (topmost branch in Fig.\ref{fig:overview}) for classification into one of the keywords. This directly models the probability $\P{c_n | K=1, S = 1 , x }$. Similarly, to model the probability $\P{K = 1 | S = 1, x}$, we pass the embeddings corresponding to the inputs containing speech through the keyword-like branch (middle branch in Fig.\ref{fig:overview}). Finally, all of the embeddings are passed through the speech branch (bottom branch in Fig.\ref{fig:overview}) to model the probability  $\P{S=1 | x}$. The outputs of the three branches are then used to compute three losses: softmax loss for keyword classification, and weighted focal loss for binary classification at the keyword and speech branches \cite{lin2017focal}. The final loss function is given as 
\vspace{-1mm}
\begin{equation}
    \calL = \calL_{\text{softmax}} + \lambda_1 \calL_{\text{keyword branch}} + \lambda_2 \calL_{\text{speech branch}},
\vspace{-1mm}
\end{equation}
where $\lambda_1, \ \lambda_2$ denote relative weights for loss function components. The different classes in each of these loss functions are weighted inversely proportional to the number of samples in the class. During backpropagation, each branch recieves gradients only from its loss component (e.g., topmost branch receives gradients from $\calL_{\text{softmax}}$ only). Only the backbone model receives gradients from all three loss components.

\vspace{0.15cm}

\noindent\textbf{Performance boost due to data pooling:} The hierarchical structure in our proposed model naturally leads to data pooling at the speech and keyword branches. For the speech branch, all the keyword and non-keyword speech audio get pooled into one class, thus boosting the size of the training data. This also allows us to use a larger non-speech dataset without creating class imbalance. Similarly, data corresponding to the various keywords also get pooled into a single class at the keyword branch. 

\vspace{0.15cm}

\noindent\textbf{Inference:} During inference of audio $x$, the keyword classification branch provides a multinomial distribution over the set of keywords $\{1,\dots,N\}$ of the form $p(c_n|K=1,S=1,x) \defeq p_{c_n}$,  the keyword branch gives a distribution whether the audio is keyword-like or not $p(K=i|S=1,x) \defeq p_{K=i}$, and the speech-branch gives a distribution whether the audio is speech or not $p(S=i|x) \defeq p_{S=i}$,  where $i \in \{0,1\}$. We combine these to get a multinomial distribution of the form $\mathbf{p} \defeq [p_{c_1}p_{K=1}p_{S=1},\dots,p_{c_N}p_{K=1}p_{S=1},p_{K=0}p_{S=1},p_{S=0}]$. Using $\mathbf{p}$ we compute $\hat{l} = \argmax_{l \in \{1,\dots,N+2\}} \mathbf{p}[l]$. If $\hat{l}=n$, where $n \in \{1,\dots,N\}$, then the $n^{th}$ keyword is detected. If $\hat{l}=N+1$ then non-keyword speech is detected and if $\hat{l}=N+2$ then non-speech is detected.
\vspace{-2mm}


%% file: Contents/Experiments.tex
\section{Experimental Results}
\label{sec:experiments}




\begin{figure*}[t!]
\begin{minipage}[b]{0.49\linewidth}
\includegraphics[height=1.8in , width=0.95\linewidth]{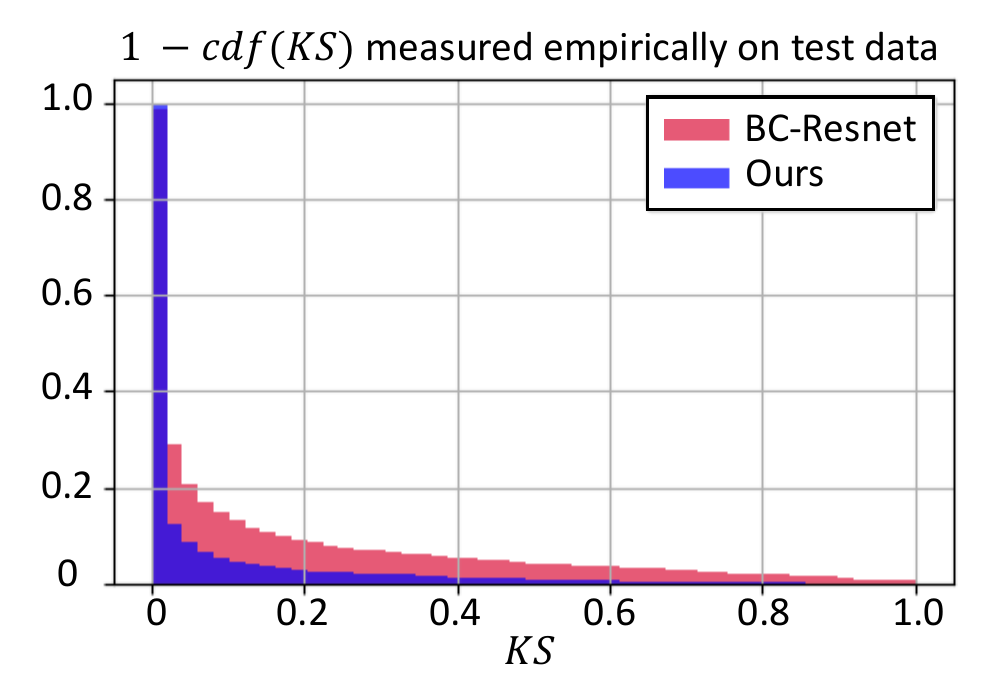}
\end{minipage}
\begin{minipage}[b]{0.49\linewidth}
\includegraphics[height=1.8in ,width=0.95\linewidth]{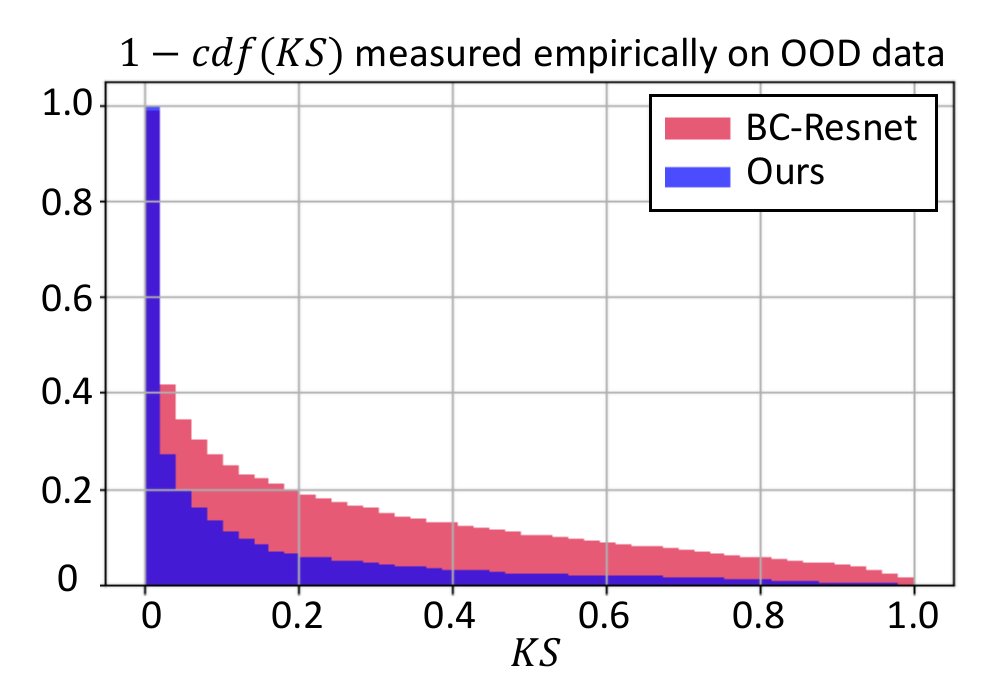}
\end{minipage}
\caption{We illustrate the inverse cumulative distribution function (cdf) of the random variable $KS$ for the non-keyword and non-speech portion of the held-out test data (left) and the OOD data (right) on BCResNet-2 with and without \textit{Succesive Refinement}. For low FA, the distribution of $KS$ should concentrate towards 0. In the above figures, we observe that $p_{\text{baseline}}(KS < \alpha) > p_{\text{ours}}(KS < \alpha) $ for most of $\alpha \in [0,1]$ for both the held-out test data and the OOD data. The effect is more pronounced on the OOD data, owing to the stronger discriminating capability of successive refinement.}
\label{fig:CDF}
\end{figure*}



\begin{figure*}[!htb]
\minipage{0.32\textwidth}
\includegraphics[width=0.9\linewidth]{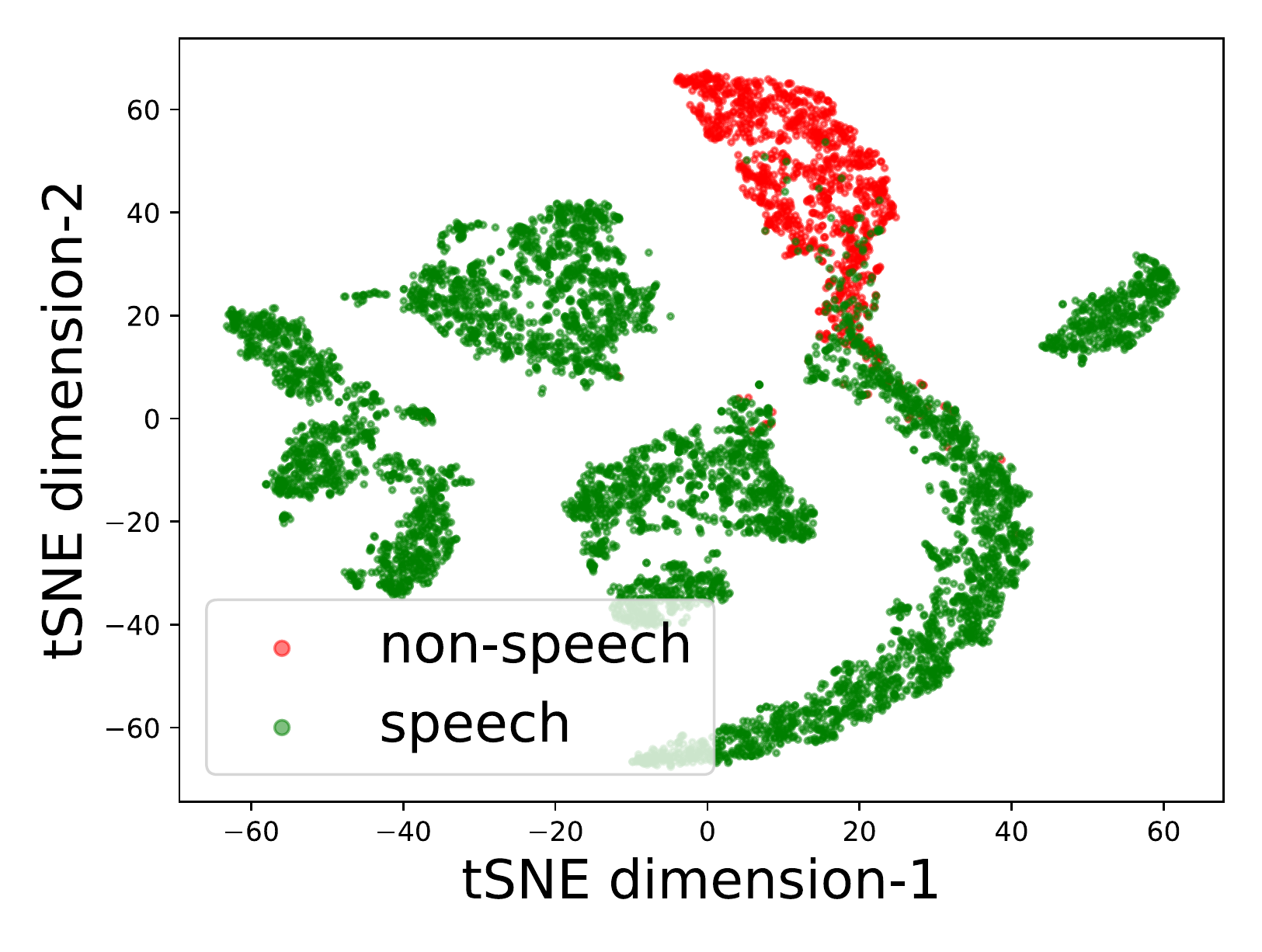}
\centering
\centerline{(a) Speech vs. non-speech branch}\medskip
\endminipage\hfill
\minipage{0.32\textwidth}
\includegraphics[width=0.9\linewidth]{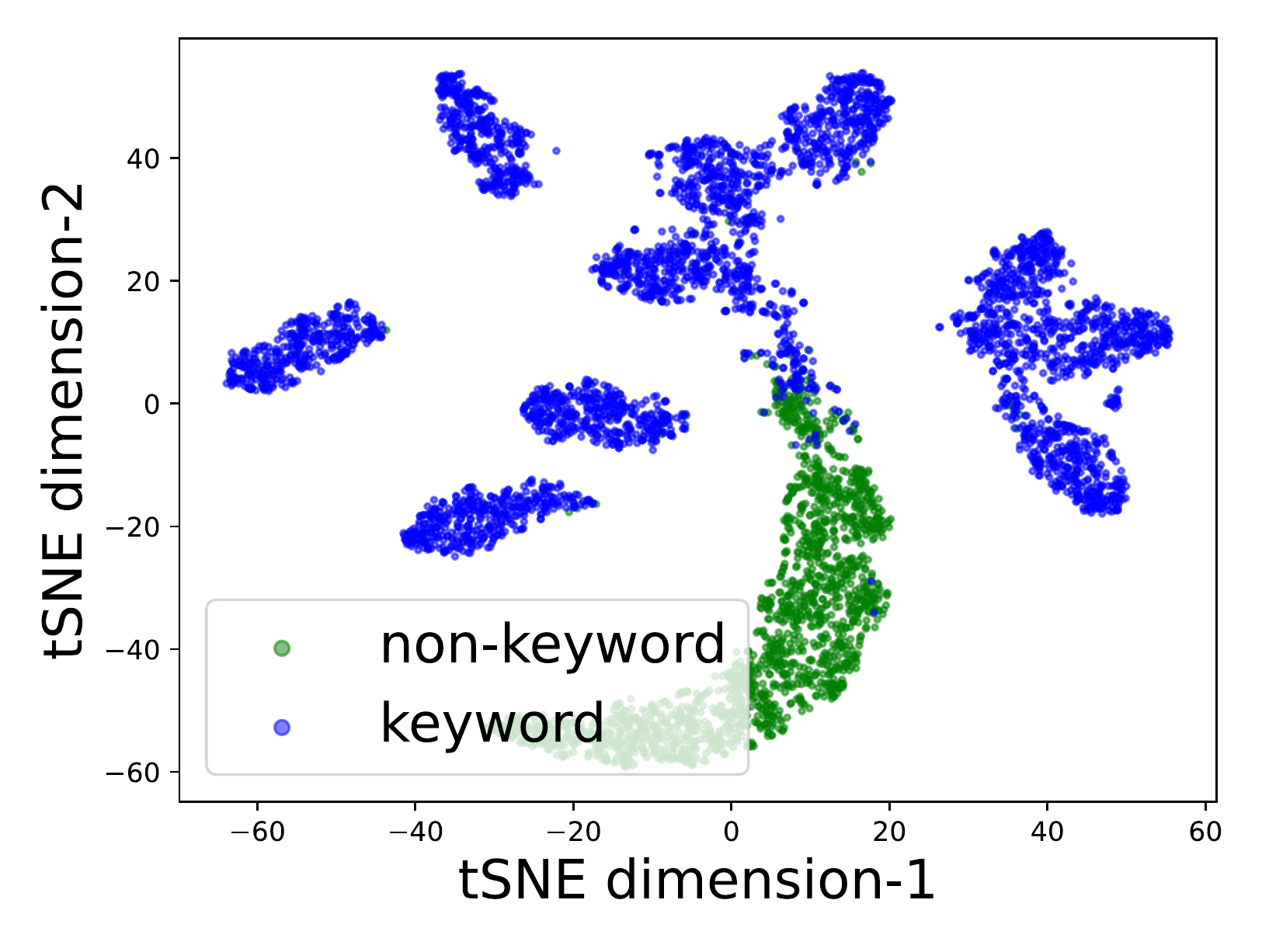}
\centering
\centerline{(b) Keyword vs. non-keyword branch}\medskip
\endminipage\hfill
\minipage{0.32\textwidth}%
\includegraphics[width=0.9\linewidth]{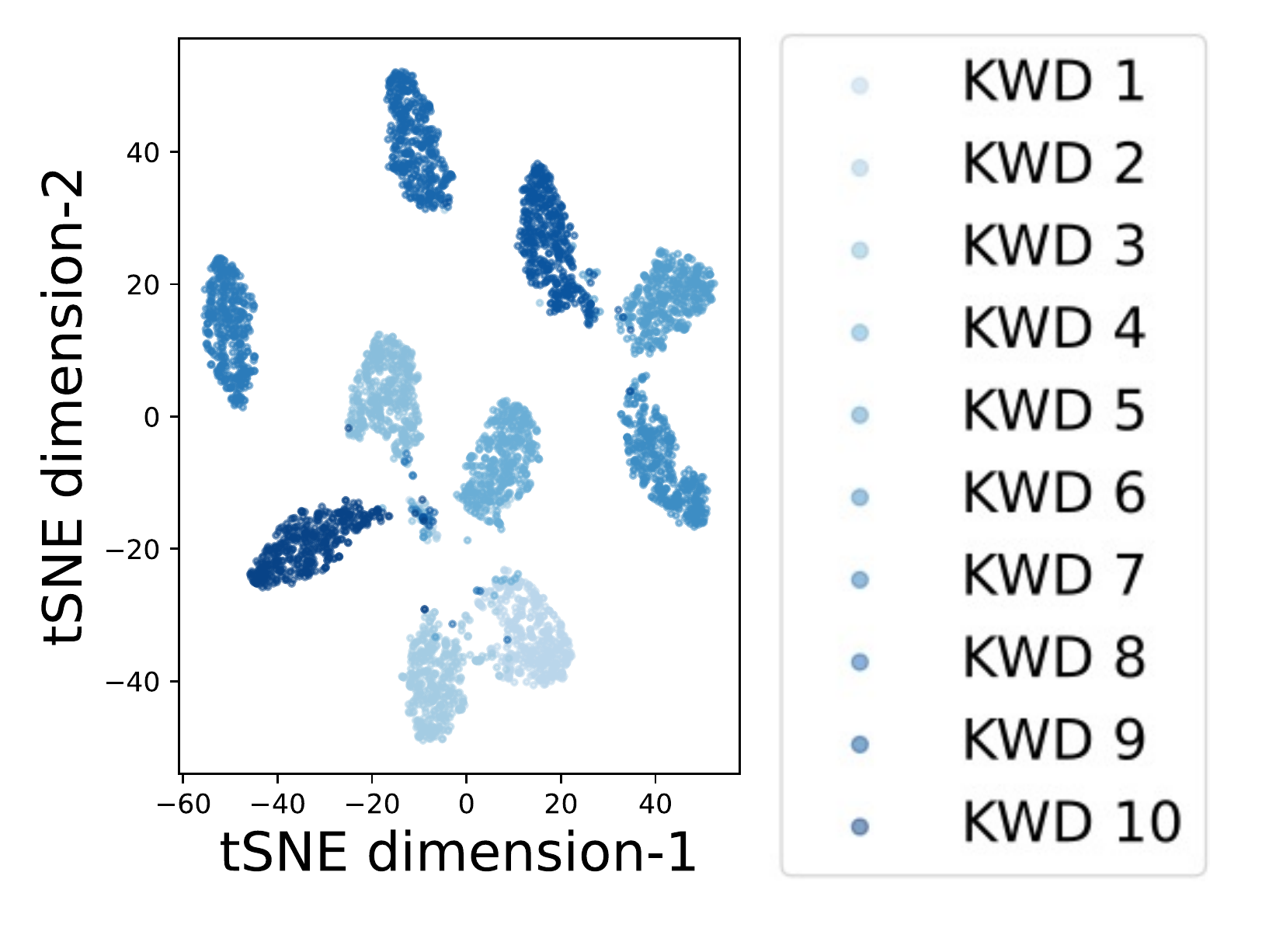}
\centerline{(c) Keyword classifier for all keywords}\medskip
\endminipage
\caption{Visualizing the embeddings learned by the three different branches using tSNE \cite{van2008visualizing}.}
\label{fig:umap}
\vspace{-4mm}
\end{figure*}


\textbf{Datasets:} We evaulate the performance of the proposed methods over the Google Speech Commands dataset V2 (GSC-V2) \cite{warden2018speech}. 
We experiment over two variants: one where only ten keywords as in \cite{warden2018speech} are used, and another where all thirty five are used. We use the same training, validation, and testing split for these classes as provided in \cite{warden2018speech}.
For the ``non-keyword Speech" (called ``Unknown" in \cite{warden2018speech}) class we use 760 one-second samples from the childspeech dataset \cite{kennedy2017child}, and 4565 one-second samples from the AVSpeech dataset \cite{ephrat2018looking}. For the ``non-speech" class (called ``Silence" in \cite{warden2018speech}) class we use 7493 one-second samples from the Audioset \cite{gemmeke2017audio}. 
Since some samples in Audioset contain speech, we filter them out using a Voice-Activity Detector \cite{sileroVAD}. All of these are split into training, validation, and testing with a 8:1:1 split.

We also perform Out-Of-Domain (OOD) experiments on the trained models to test robustness for non-keyword speech and non-speech in unseen conditions. For this, we use the MUSAN dataset \cite{snyder2015musan}. Each speech file is sampled to obtain three 1s frames and each noise file is sampled to obtain one 1s frame, resulting in a total of 2155 samples for FA testing.

\vspace{0.15cm}

\noindent \textbf{Implementation details:} We experiment on various models like the BC-ResNet models\cite{kim2021broadcasted}, Keyword Transformer\cite{berg21_interspeech}, and Time Delay Neural Network with Shared-Weight Self-Attention (TDNN-SWSA, also referred to as TSWSA for short)\cite{Bai2019time}. Our motivation in choosing these models is to consider both SOTA models (\cite{kim2021broadcasted, berg21_interspeech}) as well as small sized models (\cite{Bai2019time}).

\vspace{0.1cm}

\noindent \textbf{\textit{BC-ResNet:}} We modified the original architecture by adding an $80$ neuron linear layer followed by ReLU before the final softmax layer \cite{kim2021broadcasted}. To perform successive refinement, we replace the $80$-neuron layer with three $32$-neuron layers (one for each branch) and their own output layers. We train the models for 25 epochs using SGD and the one-cycle learning rate scheduler \cite{smith2019super}. The learning rate goes from $0.004$ to $0.1$ over seven epochs and decays to $4\times 10^{-6}$. Mini-batch size is $100$. The momentum is varied from $0.85$ to $0.95$ as per \cite{smith2019super}. All other settings like augmentation follow from \cite{kim2021broadcasted}.

\vspace{0.1cm}

\noindent \textbf{\textit{Keyword Transformer (KWT):}} For our baseline KWT models, we follow the same architecture as specified in \cite{berg21_interspeech}. We also use the same set of data augmentation described in \cite{berg21_interspeech}. To incorporate the proposed approach, we use the output of the penultimate layer. The keyword classifier consists of layer normalization followed by a single linear layer (same as in \cite{berg21_interspeech}). The speech classifier and the keyword-like classifier each consist of a linear layer of size $n \times n$ where $n$ is the embedding size, followed by ReLU and then a linear layer of size $n \times 1$. We use $N$-class cross-entropy loss for the keyword classifier and binary cross-entropy loss (after a sigmoid non-linearity) for the other two branches. We provide results only for KWT-1 and KWT-2 models, as the baseline model for KWT-3 did not converge well during training. Our implementation code was obtained from the Pytorch implementation provided at \url{https://github.com/ID56/Torch-KWT}.

\vspace{0.1cm}

\noindent \textbf{\textit{TSWSA}}: We modified the original architecture by adding a linear layer with 32 neurons followed by a ReLU before the final softmax output layer \cite{Bai2019time}. To perform successive refinement, three linear layers of 10 neurons each (and output layers) were added. The model optimizer and scheduler is the same as described for BC-ResNet. We found that a mini-batch of $100$ performed better than the original $32$. All other settings, like MFCC, follow \cite{Bai2019time}.


\vspace{0.15cm}

\noindent \textbf{Performance metrics:} We compute accuracy and weighted F1 score as a measure of overall KWS performance. This is computed across $N+2$ classes comprising of $N$ keywords, non-keyword speech, and non-speech classes. We compute FA as the fraction of non-keyword speech and non-speech audio that is classified as a keyword.

\subsection{Results}

    


We compare the improvement provided by our method (\textit{Successive Refinement} refered to as \textit{SR}) in terms of accuracy of classification, F1 score, and FA rate over in-domain unseen FA data and OOD FA data. Table~\ref{table:gsc10} shows the results where there are ten keywords. From the results we can see that \textbf{adding successive refinement consistently improves FA performance of all models over both in-domain and OOD FA data} while maintaining close to the same or better accuracy compared to baseline models. Further, \textbf{the improvement in FA is obtained without drastically increasing memory (parameter size) or computation (MACs)}. In Table~\ref{table:gsc35} we experiment on the thirty five keyword data with the BC-ResNet models, which performed the best over the ten keyword data. We find that \textbf{successive refinement has a greater impact on FA when the number of keywords is larger}.

\begin{table}[!th]
\caption{Results over training on the GSC-V2 dataset with ten keywords, a non-keyword speech class (from AVSpeech and Childspeech), and a non-speech class (from Audioset). FA(OOD) was computed using speech and noise from the MUSAN dataset. \textit{SR} refers to the proposed method of Successive Refinement.}
\label{table:gsc10}
\vspace{-5mm}
\begin{center}
\begin{footnotesize}
\setlength{\tabcolsep}{3pt} 
\begin{tabular}{l|cccccc}
\toprule
 Model & Acc (\%) & F1 & FA (\%) & \begin{tabular}{@{}c@{}}FA OOD \\ (\%)\end{tabular} & \#Params & \#MACs\\
\midrule
 BCR-1 \cite{kim2021broadcasted} & 93.42 & 0.965 & 4.10 & 6.08 & 13.0k & 3.34M\\
 BCR-1+SR & 93.22 & 0.964 & \textbf{1.41} & \textbf{2.13} & 13.0k  & 3.34M\\
\midrule
 BCR-1.5 \cite{kim2021broadcasted} & 94.05 & 0.967 & 3.99 & 6.45 & 22.3k & 5.89M\\
 BCR-1.5+SR & 94.24 & 0.969 & \textbf{1.17} & \textbf{2.65} & 22.5k & 5.89M\\
\midrule
 BCR-2 \cite{kim2021broadcasted} & 94.17 & 0.969 & 4.23 & 8.03 & 33.8k & 9.03M\\
 BCR-2+SR & 95.26 & 0.975 & \textbf{1.11} & \textbf{2.22} & 34.3k & 9.03M\\
\midrule
 BCR-3 \cite{kim2021broadcasted} & 95.48 & 0.976 & 3.79 & 5.89 & 63.5k & 17.1M\\
 BCR-3+SR & 95.77 & 0.978 & \textbf{1.01} & \textbf{2.13} & 64.5k & 17.1M\\
\midrule
 BCR-6 \cite{kim2021broadcasted} & 96.48 & 0.982 & 2.25 & 4.36 & 205k & 55.3M\\
 BCR-6+SR & 96.27 & 0.981 & \textbf{1.04} & \textbf{1.58} & 208k & 55.3M\\
\midrule
 BCR-8 \cite{kim2021broadcasted} & 96.70 & 0.983 & 2.15 & 4.22 & 344k & 92.6M\\
 BCR-8+SR & 96.82 & 0.984 & \textbf{0.77} & \textbf{1.72} & 348k & 92.6M\\
\midrule 
\midrule 
KWT-1 \cite{berg21_interspeech} & 94.9 & 0.968 & 2.5 & 7.19 & 551k & 56.40M \\
KWT-2+SR & 94.48 & 0.965 & \textbf{2.3} & \textbf{5.38}& 559.7k & 56.42M \\
\midrule 
KWT-2 \cite{berg21_interspeech} & 94.62 & 0.967 & 2.75 & 5.75 & 2,381k & 243.2M \\
KWT-2+SR &94.76& 0.967 & \textbf{2.08} & \textbf{5.38} & 2,415k&243.2M \\
\midrule 
\midrule 
TSWA \cite{Bai2019time} & 91.56 & 0.955 & 4.91 & 9.84 & 12.8K & 0.36M\\
TSWSA+SR & 91.21 & 0.953 & \textbf{1.78} & \textbf{2.97} & 12.5K & 0.36M \\
\bottomrule
\end{tabular}
\end{footnotesize}
\end{center}
\vspace{-5mm}
\end{table}


\begin{table}[!th]
\caption{Results over training on the GSC-V2 dataset with thirty five keywords. \textit{SR} refers to proposed method of Successive Refinement.}
\label{table:gsc35}
\vspace{-6mm}
\begin{center}
\begin{footnotesize}
\setlength{\tabcolsep}{4pt} 
\begin{tabular}{l|cccccc}
\toprule
 Model & Acc (\%) & F1 & FA (\%) & \begin{tabular}{@{}c@{}}FA OOD \\ (\%)\end{tabular} & \#Params & \#MACs\\
\midrule
 BCR-1 \cite{kim2021broadcasted} & 90.17 & 0.944 & 13.49 & 14.06 & 15.1k & 3.34M \\
 BCR-1+SR & 89.65 & 0.942 & \textbf{1.61} & \textbf{1.95} & 13.8K & 3.34M \\
\midrule
 BCR-1.5 \cite{kim2021broadcasted} & 91.61 & 0.953 & 12.02 & 15.41 & 24.4k & 5.89M \\
 BCR-1.5+SR & 91.38 & 0.952 & \textbf{1.58} & \textbf{2.55} & 23.4K & 5.89M \\
\midrule
 BCR-2 \cite{kim2021broadcasted} & 92.58 & 0.959 & 10.47 & 13.36 & 35.9k & 9.03M \\
 BCR-2+SR & 93.00 & 0.962 & \textbf{1.98} & \textbf{3.62} & 35.1K & 9.03M \\
\midrule
 BCR-3 \cite{kim2021broadcasted} & 93.73 & 0.965 & 8.69 & 11.69 & 65.5k & 17.1M \\
 BCR-3+SR & 94.23 & 0.968 & \textbf{1.61} & \textbf{2.83} & 65.3K & 17.1M \\
\midrule
 BCR-6 \cite{kim2021broadcasted} & 95.10 & 0.973 & 6.71 & 9.47 & 207k & 55.3M \\
 BCR-6+SR & 95.54 & 0.976 & \textbf{1.21} & \textbf{1.58} & 209K & 55.3M \\
\midrule
 BCR-8 \cite{kim2021broadcasted} & 95.39 & 0.975 & 6.04 & 8.86 & 346k & 92.6M \\
 BCR-8+SR & 95.56 & 0.976 & \textbf{0.91} & \textbf{1.76} & 349K & 92.6M \\
\bottomrule
\end{tabular}
\end{footnotesize}
\end{center}
\vspace{-10mm}
\end{table}


To further study the effect of our approach on FA rates, we illustrate the cumulative distribution of $K S$ (the random variable that indicates whether a given audio input is both speech and one of the keywords) in Fig. \ref{fig:CDF}.  Note that $K S$ is the maximum possible probability for any keyword. A small upper bound on $K S$ hence demonstrates a low FA rate.  For the baseline model, we estimate $K S$ empirically as $K S = 1 -\sum_{i=1}^N p(c_i)$ where $c_i$ is the $i$th keyword. Our approach (shown in blue) always results in lower values of $K S$, leading to \textbf{a better worst-case FA rate.} This is also reflected in our FA rates shown in Tables \ref{table:gsc10} and \ref{table:gsc35}.

Our proposed approach branches the problem into first classifying whether the input audio is speech or not, if it is speech whether it is keyword-like or not, finally, if it is keyword-like which keyword was uttered. This helps the system learn specialized embeddings that are easier to perform each of these sub-tasks. In Fig.~\ref{fig:umap} we visualize these embeddings using tSNE \cite{van2008visualizing}. In Fig. \ref{fig:umap} (a), the speech branch just focuses on separating speech from non-speech instead of clustering the different speech types into different clusters. However, the following keyword branch shown in Fig. \ref{fig:umap} (b), separates the speech types into keyword-like speech and general speech. Finally, as shown in Fig. \ref{fig:umap} (c), the keyword classifier clusters the different keywords into their own clusters. Thus, \emph{Successive Refinement} lets different branches learn specialized embeddings for simpler sub-problems, resulting in better discriminative power.




%% file: Contents/Conclusion.tex
\section{Conclusion}
\label{sec: conclusion}
\vspace{-2mm}

Existing KWS systems treat non-keyword speech and non-speech as two classes equally far-apart from the keyword classes. However, in reality non-keyword speech and keywords are more similar to each other than to non-speech. We exploit this insight to build a successive refinement classifier that classifies whether the input audio is speech or not speech, if speech whether it is keyword-like or generic speech, and finally if it is keyword-like which keyword was uttered. This also follows as a simple, yet elegant expansion of the law of total probability. We conduct experiments across models of various sizes from 13K parameters to 2.41M parameters and show that the detection accuracy of the model is similar to the baseline while the FAs are reduced by up to a factor of 8 on in domain FA data and up to a factor of 7 on OOD FA data.